\documentclass{jpp}
\usepackage{graphicx}
\usepackage{epstopdf, epsfig}
\usepackage{hyperref}
\usepackage[utf8]{inputenc}
\usepackage{amsmath,amsfonts,amssymb}
\usepackage{natbib}
\usepackage{multirow}

\DeclareUnicodeCharacter{2009}{\,} 
\usepackage[LGRgreek]{mathastext}

\shorttitle{EP LILA }
\shortauthor{R. S.  Rajawat, T. Wang and G. Shvets }

\title{Effects of Laser Polarization on Target Focusing and Acceleration in a Laser-Ion Lens and Accelerator (LILA)}

\author{Roopendra Singh Rajawat\aff{1} \corresp{\email{rupn999@gmail.com}}, Tianhong Wang\aff{1},  V. Khudik\aff{2} and  Gennady Shvets\aff{1}}

\affiliation{\aff{1} School of Applied and Engineering Physics, Clark Hall, Cornell University, Ithaca, NY, USA-14853 \aff{2} Department of Physics and Institute for Fusion studies, The University of Texas at Austin, TX 78712}

\begin{document}

\maketitle

\begin{abstract}
We present the process of ion acceleration using ultra-thin foils irradiated by elliptically polarized, high-intensity laser pulses. Recently, efficient generation of monoenergetic ion beams was introduced using the concept of laser-ion lensing and acceleration (LILA) \citep{Wang2021}.  LILA is an innovative technique where the target’s radially varying thickness enables simultaneous acceleration and focusing of a proton beam. In this work, we extend the LILA framework to incorporate elliptically polarized (EP) laser pulses. While it's commonly assumed that EP lasers are unsuitable for radiation pressure acceleration (RPA) due to excessive electron heating that compromises ion acceleration, our multidimensional particle-in-cell simulations challenge this notion. We show that, with proper optimization of the target's average thickness, EP laser pulses can successfully drive the LILA mechanism. We also demonstrate that with a non-uniform thickness target, even linearly polarized laser pulses can efficiently generate low-emittance focused ion beams,  with the overall laser-to-ions energy conversion comparable to those predicted for circularly polarized laser pulses.
 \end{abstract}

\section{Introduction}
The generation of compact ion beams from laser- plasma interactions has garnered significant interest due to their low emittance and monoenergetic energy peak. These ion beams have a range of applications, including fast ignition of fusion targets \citep{Tabak1994,Atzeni2002}, hadron cancer therapy \citep{Dyer2008,Patel2003,Kroll2022}, and particle nuclear physics \citep{Bulanov2002,Yogo2009,Bolton2010,Hannachi2007}. 

These ion beams are generated using two widely popular schemes: Radiation Pressure Acceleration (RPA) \citep{Esirkepov2004,Esirkepov2005,Robinson2008,Qiao2009,Wang2021,Kim2023} and Target Normal Sheath Acceleration (TNSA) \citep{Wilks_TNSA,Pukhov_PRX}. In the RPA scheme, an electron sheath is driven forward by laser radiation pressure, inducing a charge-separation electrostatic field that drags and accelerates ions in the direction of the electron sheath. In TNSA, laser-heated electrons escape through the rear surface, generating an electrostatic sheath field that accelerates ions from the rear surface of the target.
Both schemes act on opposite surfaces of a target and may function either independently \citep{Kim2016} or synergistically \citep{Qiao2009,Ziegler2024}, depending on laser intensity and target geometry. A trivial target shape, \textit{viz.} flat foil, is highly susceptible to the laser’s transverse spot size \citep{Dollar2012} and polarization \citep{Bulanov_prl_2015}. The transverse spot size can cause the thin foil to expand in the transverse dimension, leading to relativistic self-induced transparency (RSIT) \citep{Qiao2009,Kar2012,Scullion2017}. While RSIT can accelerate ions to high energies \citep{Ziegler2024,Gonzalez_np_2016,Gonzalez_nc_2016}, it relies heavily on selective ion acceleration. Instead of accelerating the entire target foil, only a small overdense electron layer is accelerated via radiation pressure, which results in a localized ion layer with a broad, non-monoenergetic energy distribution. Therefore, it is crucial for the target to remain opaque during the laser-target interaction. Another challenge with a trivial target shape is the onset of instabilities, such as Weibel and Rayleigh-Taylor-like instabilities \citep{Palmer2012,Khudik2014}, which breaks the front surface and reduce ion acceleration efficiency. Some of these effects can be mitigated by shaping the target such that target density increases with acceleration, thereby preventing self-induced transparency.

Based on TNSA, several studies have proposed ion focusing using a plasma-based microlens \cite{Toncian_Science_2006} or by shaping a thin foil into a hemisphere \citep{Patel2003,Snavely_pop_2007,Offermann_pop_2011} with a guiding cone positioned behind it \citep{Bartal_nphys_2012, Qiao_pre_2013,McGuffey_sr_2020}. Although TNSA is not significantly affected by Rayleigh-Taylor-like instabilities, it is less efficient for accelerating heavy ions and often requires post-acceleration energy selection to achieve the desired emittance, as its broad (exponential) energy spectrum can reduce overall efficiency. 

\begin{figure} \label{Fig1}
\centering
\includegraphics[width=0.8\linewidth]{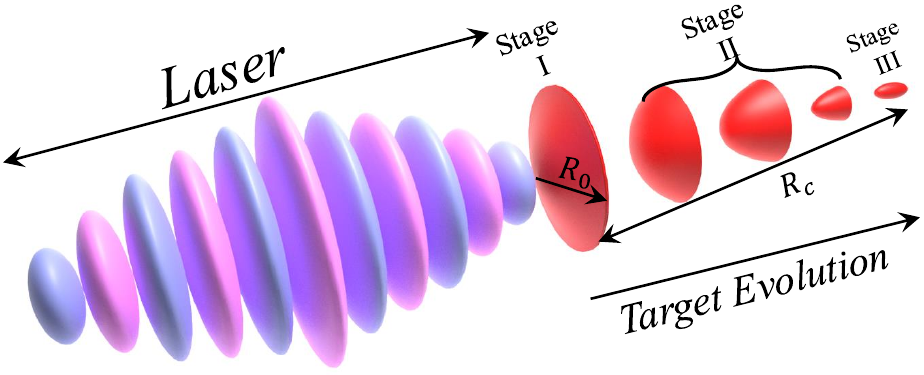}
\caption{The schematic of the Laser-Ion Lens and Accelerator (LILA) illustrates the different stages of the process. Stage II represents the phase when the laser pulse begins to deform the target, causing it to bend forward as the laser radiation pressure builds. During this stage, the interaction between the laser and the target leads to ion acceleration as the target responds to the intense electromagnetic fields. Stage III occurs after the laser-target interaction has concluded. At this point, the target focuses, achieving maximum ion density as the ions are collimated into a compact beam.}
\end{figure}

Recently, the \textit{Laser-Ion Lens and Acceleration} (LILA) concept, based on light-sail radiation pressure acceleration, has been introduced to produce collimated, high-flux, high-energy ion bunches \citep{Wang2021}. LILA operates similarly to an optical lens; however, instead of focusing light, it focuses a matter target shaped like a lens to a dense focal point using intense photon radiation pressure.  Figure \ref{Fig1} illustrates the interaction of a laser with a thin overdense target, whose thickness decreases from the center toward its edges in a parabolic profile (stage I).  For a Gaussian laser pulse interacting with a parabolic target of radius $R_0$, the radially dependent thickness, $d(r_0)$, of the target is defined as: $d(r_0) =  d_0 exp(-r^2/R_c^2) exp(-r^2/\sigma_L^2)$, where $r$ is the transverse coordinate, $d_0$ is the target thickness at its center, $R_c$ is the radius of curvature, and $\sigma_L$ is the size of the laser's focal spot. Due to this shape, the outer (thinner) regions of the target accelerate faster than the central (thicker) region, causing the initially flat target to curve forward (stage II; Fig. \ref{Fig1}) and focus into a small volume (stage III; Fig. \ref{Fig1}) at a designed focal length $R_c$. By simultaneously focusing and accelerating ions, the circularly polarized LILA target remains resistant to induced relativistic transparency over a wide parameter range.

The effect of the laser-polarization on RPA is dramatic and important to understand because radial distribution of accelerated electrons has strong dependence on laser polarization \citep{Gonzalez_nc_2016}. The ponderomotive force when a linearly polarized laser pulse irradiates a solid thin target is given by \citep{Rykovanov2008}

\begin{equation} \label{eq:eq1}
f_p(x) = - \frac{e^2}{4 m_e \omega_l^2} \nabla \left\langle E_l^2 \right\rangle \left(1 - \left( \frac{1 - \epsilon^2}{1+\epsilon^2} \right) \cos \left(2 \omega t\right) \right)
\end{equation} 

where $e$ is the electron charge, $m_e$ the electron rest mass, $\omega$ the laser angular frequency and $\epsilon = E_y/E_z$ is the laser polarization ellipticity ($0 \le \epsilon \le  1$); where $E_y$ and $E_z$ are the electric field polarized in $y$ and $z-$directions, respectively. Note that at relativistic intensities the force on the electrons arising from $\bf J \times B$ term in the Lorentz equation is of the same order as that due to the electric field. The first term on the right hand side of ponderomotive force (equation \eqref{eq:eq1}) pushes electrons from higher to lower electric fields. The second term is the $\bf J \times B$ heating mechanism \citep{Brunel1987,Brunel1988} and excites electrons oscillation at twice the laser frequency ($2 \omega$). For circular polarization $\epsilon = 1$, the $\bf J \times B$ heating component of equation \eqref{eq:eq1} becomes zero, consequently, electrons are adiabatically pushed by radiation pressure as a single layer and ions are effectively accelerated in the non-oscillating charge separating field. For linear polarized pulses $\epsilon = 0$, the ellipticity factor $(1-\epsilon^2)/(1+\epsilon^2)$ becomes 1. Unlike CP, in this case, oscillating component of $\bf J \times B$ force drives the electrons in longitudinal motion back and forth through the thin target foil. Thus, electrons are heated by $\bf J \times B$ component and leads to foil explosion rather than smooth localized ion acceleration. Laser polarization is thus highly important in defining the coupling of laser energy to target plasma electrons at relativistic laser intensities. Effect of laser polarization on radiation pressure acceleration with an overdense thin planar foil target has been studied and reported generation of hot electrons and exponential ion energy spectrum \citep{Liseykina_ppcf_2008,Henig_prl_2009}. We conclude that interaction of a linearly polarized laser pulse with a flatfoil target produces more energetic ions than a circularly polarized laser, albeit it has large emittance and broad ion energy spectrum. To the fore post-accceleration of selective energy selection is still required.

Intuitively we ask the following question : can a laser-ion lens and accelerator generate low-emittance, monoenergetic compact ion bunches using elliptically and linearly polarized laser pulses? In this paper, we aim to address this question. Using ab-initio 3D first-principles particle-in-cell simulations, we present that LILA works efficiently when target thickness is optimized $d_{opt} = (\lambda_l/\pi)(n_c/n)a_0$ \citep{Machi_PRL_2009,Henig_prl_2009,Kim2016} for a peak laser electric field ($a_0 = eE/m_e \omega_l c$) to avoid induced relativistic transparency \citep{Vashikov_pop_1998}, where $c$ being the speed of light. We demonstrate that the LILA target can generate collimated, high-flux monoenergetic ion bunches regardless of the laser pulse polarization. Additionally, we show that LILA can produce ion beams with energies comparable to those generated using circularly polarized lasers. We discuss 6 cases based on laser polarizations (3 types) and target shapes (2 types) of the target. The cases are following: (I) shaped target; CP pulse ($\epsilon = 1$), (II) planar target; CP pulse ($\epsilon = 1$), (III) shaped target; EP pulse ($\epsilon = 0.5$), (IV) planar target; EP pulse ($\epsilon = 0.5$), (V) shaped target; LP pulse ($\epsilon = 0$), (VI) planar target; LP pulse ($\epsilon = 0$). In Secion II we discuss cases I and II, in secion III we discuss cases III and IV, and in section IV we discuss cases V and VI. A comparative discussion is done in section V. We present conclusion in section VI.

\begin{center}
\begin{tabular}{|p{3cm}|p{3cm}|p{3cm}|p{3cm}|} 
\hline
 Case & Target Geometry &$\epsilon$ & $d_0 (nm)$  \\ [0.5ex] 
 \hline\hline
 1 & Shaped & 1 & 0.21 \\ 
 2 & Planar & 1 & 0.13 \\
 3 & Shaped & 0.5 & 0.27 \\ 
 4 & Planar & 0.5 & 0.17 \\
 5 & Shaped & 0 & 0.29 \\ 
 6 & Planar & 0 & 0.18 \\
 \hline
\end{tabular}
\end{center}

We use 3D first-principles particle-in-cell code (VLPL) \citep{Pukhov_jpp_1999}. The box length is $L_x  = 20 \lambda_l$ in laser propagation direction from $x = -5 \lambda_l$ to $x = 15 \lambda_l$, and $L_y = L_z = 12 \lambda_l$ in transverse direction. The simulation box contains 150 cell in transverse direction and 2000 cells in $x-$direction. The simulation box has periodic boundaries in transverse direction and absorbing in laser propagation (x-) direction.

\begin{figure}
\includegraphics[width=1\linewidth]{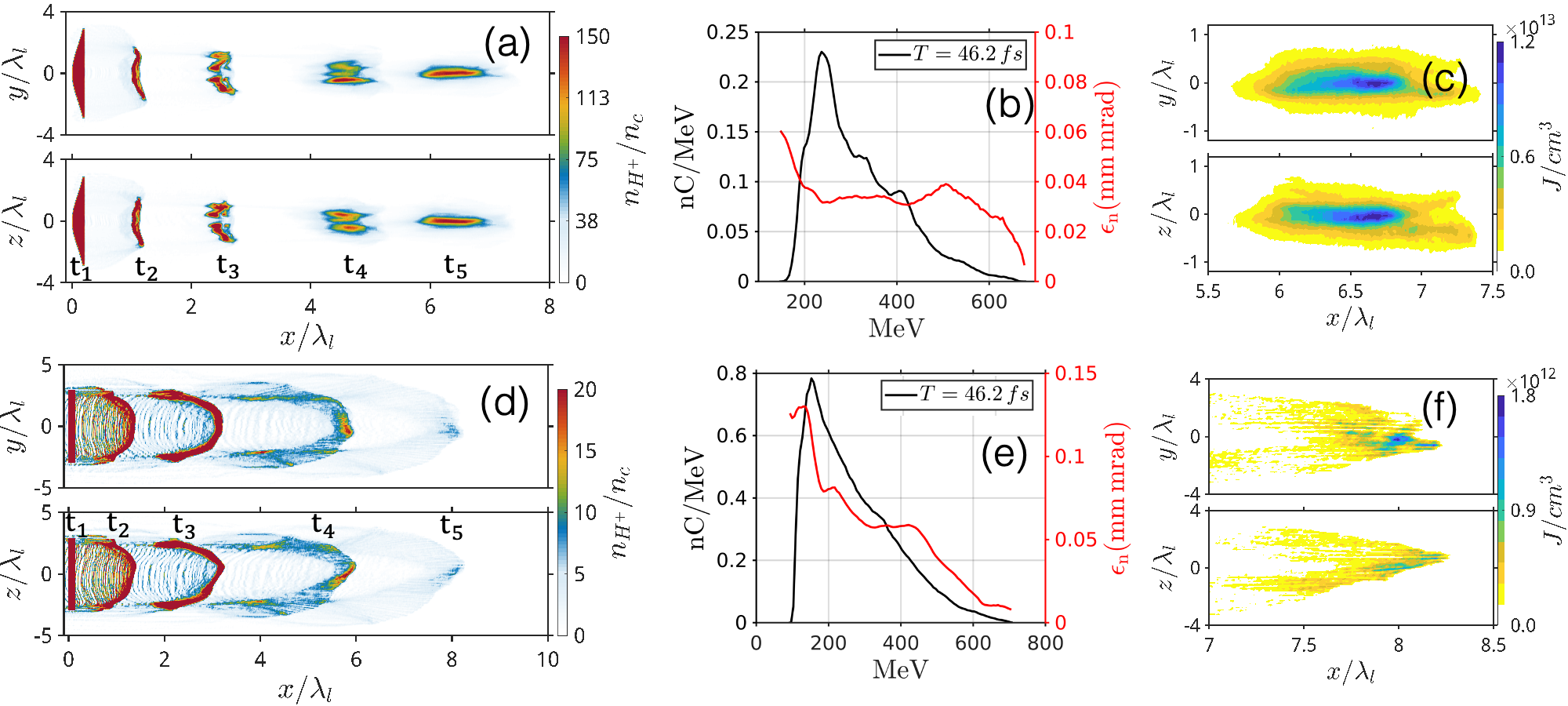} 
\caption{Ion density, energy spectrum and energy density. All figures are plotted for CP laser pulse with $a_0 = 80$.  Top Panel: Shaped Target.  (a) Time snapshots of ion density at $\mathrm{t_1 = 0} $, $\mathrm{t_2 = 13.2 \, fs} $, $\mathrm{t_3= 13.7 \, fs}$, $\mathrm{t_4 = 35.6 \, fs}$, and $\mathrm{t_5 = t_f= 46.2 \, fs} $. Ion density peaks at $t = 46.2 \, fs$.(b) Proton energy spectrum (black line; left scale) and emittance (red line; right scale) at the time of density peak.  (c) Snapshot of proton energy density at the time of density peak. Bottom Panel: Planar Target.(d) Time snapshots of ion density at $\mathrm{t_1 = 0} $, $\mathrm{t_2 = 13.2 \, fs} $, $\mathrm{t_3= 13.7 \, fs}$, $\mathrm{t_4 = 35.6 \, fs}$, and $\mathrm{t_5 = t_f= 46.2 \, fs} $. (e) Ion energy spectrum and (f) Snapshot of Proton energy density at the time $t = 46.2 \, fs$.}
\label{Fig2}
\end{figure}

 The fully ionized proton plasma foil of initial density $n_{H^{+},e} = 150 n_c$ is kept at $x = 0 $, where $n_c$ is critical density. The thickness at the center of the target is $d_0 = 0.21 \mu m$, effective radius of curvature $R_c = 2.4 \mu m$ and radius of the target is $R_0 = 2.8 \mu m$.  The laser pulse has a duration of $7 \tau_0 $, consisting of a plateau region of $5 \tau_0$ and rising and falling times of $1 \tau_0$ each, where $\tau_0 = 3.3 fs$. The intensity of the laser is $I_0 = 1.7 \times 10^{22} \, W/cm^2 $, which yields  peak electric field $a_0$ = 80 for the CP laser with the target radius $R_0$. To keep the intensity constant for different polarized laser pulse, we increase peak electric field from $a_0 = 80$  to $a_0 = 112$ for LP laser pulses. These parameters correspond to the estimated power $P \simeq 4 \, PW$, which is readily available at facility \citep{Kim2016}. The shaped target areal mass density is kept same by choosing appropriate thickness for planar target. The radius remains same for all target. The radius of curvature and total radius is identical for all simulation runs for shaped targets, and the total radius is identical for all shaped and planar targets unless stated otherwise. 

\section{Effect of a circularly polarized laser}
To establish a baseline for comparison, we first present results of laser-target interactions using a circularly polarized (CP) laser pulse. Figure~\ref{Fig2}(a-c) illustrates the results for a shaped target (case I) and a planar target (case II, Fig.~\ref{Fig2}(d-f)) under CP laser illumination, with a peak normalized vector potential of $a_0 = 80$. The central thickness of the shaped target is $d_0 = 0.21 \mu m$, while for the planar target, $d_0 = 0.13 \mu m$, adjusted to maintain an equivalent mass density between both target types. The curvature radius and total radius are as previously defined.

The spatio-temporal evolution of the shaped (Fig.\ref{Fig2}(a)) and planar (Fig.\ref{Fig2}(d)) targets demonstrate the differences in behavior under CP laser interaction. In both cases, the target is accelerated by the non-oscillating component of the $\bf J \times B$ force of the CP pulse. Notably, the shaped target achieves focusing at an effective focal length $R_c$ and retains an ion density near its initial value, $n_{H^{+}} = 150 n_c$, at the time of peak compression ($t_f = 46.2 fs$). In contrast, the planar target undergoes transverse expansion due to the finite width of the laser pulse \citep{Dollar2012}, resulting in a reduced density of approximately $n_{H^+}/n_c \sim 20$.

The proton energy spectrum (black line, left axis) and normalized emittance (red line, right axis) are presented in Fig.\ref{Fig2}(b) for the shaped target and in Fig.\ref{Fig2}(e) for the planar target. For the shaped target, the proton energy peaks at $E_k \approx 230 \, MeV$ with a very low normalized emittance of $\epsilon_n \approx 0.032 \,  mm \, mrad$ at the peak of the energy spectrum. In comparison, the planar target, shown in Fig.~\ref{Fig2}(d), exhibits Rayleigh-Taylor instability and significant transverse expansion, leading to a lower energy peak at $E_k \approx 190 \, MeV$ and a larger normalized emittance of $\epsilon_n \approx 0.1 \, mm \, mrad$.

Figure~\ref{Fig2}(c) displays the proton energy density at the target focusing time, $t_f$, peaking at approximately $1.2 \times 10^{13} \, J/cm^3$ for the shaped target, while the planar target achieves a much lower peak energy density of $1.8 \times 10^{12} \, J/cm^3$, nearly an order of magnitude less. These results underscore the advantages of using a shaped target, even with moderate-power laser pulses.

\section{Effect of an ellipticaly polarized laser}

\begin{figure}[t]
\includegraphics[width=1\linewidth]{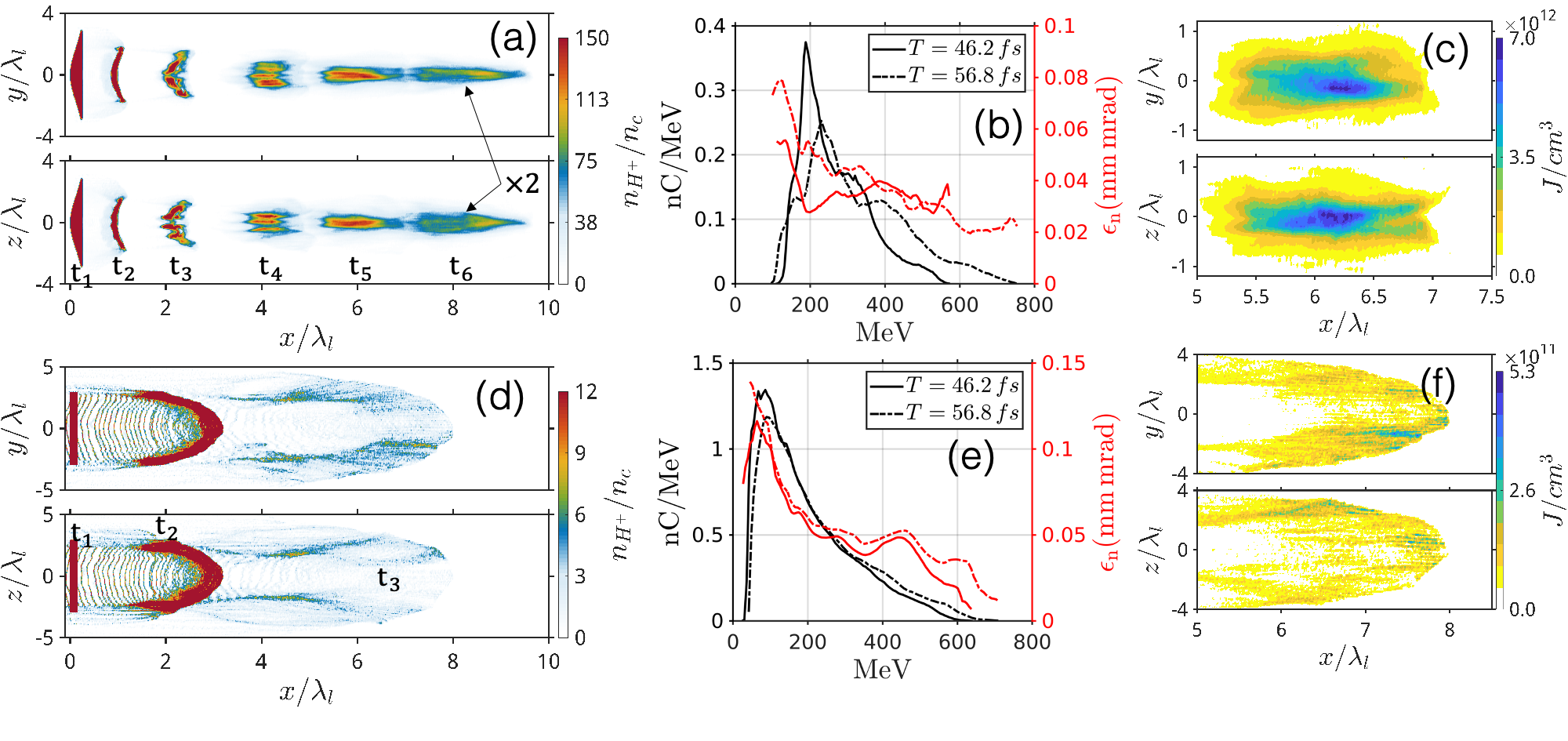} 
\caption{All figures are plotted for EP laser pulse with $\epsilon = 0.5; a_0 = 101$.  Top Panel: Shaped Target.  (a) Time snapshots of ion density at $\mathrm{t_1 = 0} $, $\mathrm{t_2 = 13.2 \, fs} $, $\mathrm{t_3= 13.7 \, fs}$, $\mathrm{t_4 = 35.6 \, fs}$,  $\mathrm{t_5 = t_f= 46.2 \, fs}$, and $\mathrm{t_6 = 56.8 \, fs} $.  Ion density peaks at $t = 46.2 \, fs$.(b) Proton energy spectrum (black line; left scale) and emittance (red; right scale) at time instances $t = 46.2 \,  fs$ (solid) and $t = 56.8 \,  fs$ (dash-dot).  (c) Snapshot of proton energy density at the time of density peak. Bottom Panel: Planar Target.(d) Time snapshots of ion density at $\mathrm{t_1 = 0} $, $\mathrm{t_2 = 23.7 \, fs} $,  and $\mathrm{t_3= 46.2 \, fs}$.  (e) and (f) follow (b) and (c), respectively,  for the planar target.}
\label{Fig3}
\end{figure}

In Fig.\ref{Fig3}(a-c), we present the results for both a shaped (LILA, Fig.\ref{Fig3}(a-c)) and a planar target (Fig.~\ref{Fig3}(d-f)) under the influence of an EP laser pulse, where $\epsilon = 0.5$ and the peak normalized vector potential of the laser is $a_0 = 108$. The central thickness of the shaped target is optimized to $d_0 = 0.27 \, \mu m$ to balance the radiation pressure and space charge forces, while the planar target has a thickness of $d_0 = 0.170 \, \mu m$. Figures \ref{Fig3}(a) and \ref{Fig3}(d) illustrate the spatio-temporal evolution of the shaped and planar targets, respectively. In both cases, target acceleration is driven by the non-oscillating component of the $\bf J \times B$ force from the EP pulse, while the oscillating component rapidly heats the electrons. For the shaped target, the production of hot electrons does not significantly affect the acceleration process due to the radial focusing force, which enhances density over time and maintains target opacity, allowing the RPA mechanism to continue. This effect is evident in Fig.~\ref{Fig3}(a), where the protons reach maximum compression at $t_f$, achieving a density peak of $n_{H^{+}} \simeq 150 n_c$ at the hotspot. In contrast, the planar target experiences significant disruption from hot electron generation and radial expansion due to the finite laser pulse width, as shown in Fig.\ref{Fig3}(d) at $t = 23.7 \, \mathrm{fs}$. Additionally, the onset of Rayleigh-Taylor (RT) instability impacts the front surface of the planar target. Consequently, the planar target undergoes early-stage transparency, as observed in Fig.\ref{Fig3}(d) at $t = 46.2 \, \mathrm{fs}$, effectively halting the RPA mechanism. For multispecies targets, heavier ions can stabilize lighter ions against RT instability \cite{Yu2011}, though this effect is not relevant for the shaped target, which remains opaque to the laser fields throughout the simulation duration. This difference is further apparent in the proton energy spectra. Figure \ref{Fig3}(b) (left scale) shows the proton energy spectrum for the shaped target, where the proton energy peaks at approximately $E_k \approx 200\, \mathrm{MeV}$, with a low normalized emittance of $\epsilon_n \simeq 0.035$ at the energy peak. Conversely, the planar target's proton energy peaks at only $E_k \approx 100 \, \mathrm{MeV}$, highlighting the adverse effects of early relativistic transparency. The proton energy density for the shaped target reaches $\sim7 \times 10^{12} \, \mathrm{J/cm^3}$, which is an order of magnitude higher than that of the planar target ($5.3 \times 10^{11} \, \mathrm{J/cm^3}$), as illustrated in Fig.~\ref{Fig3}(e).

In Fig. \ref{Fig3}(c) proton energy density is plotted at the time of target focusing $t_f$ and peaks at $\sim 1.2 \times 10^{13} J/cm^3$ while planar target energy density peaks at $\sim 1.2 \times 10^{13} J/cm^3$ (see Fig. \ref{Fig3}(f)) due to limitation by transverse expansion. 



\section{Effect of a linearly polarized laser}

\begin{figure}
\includegraphics[width=1\linewidth]{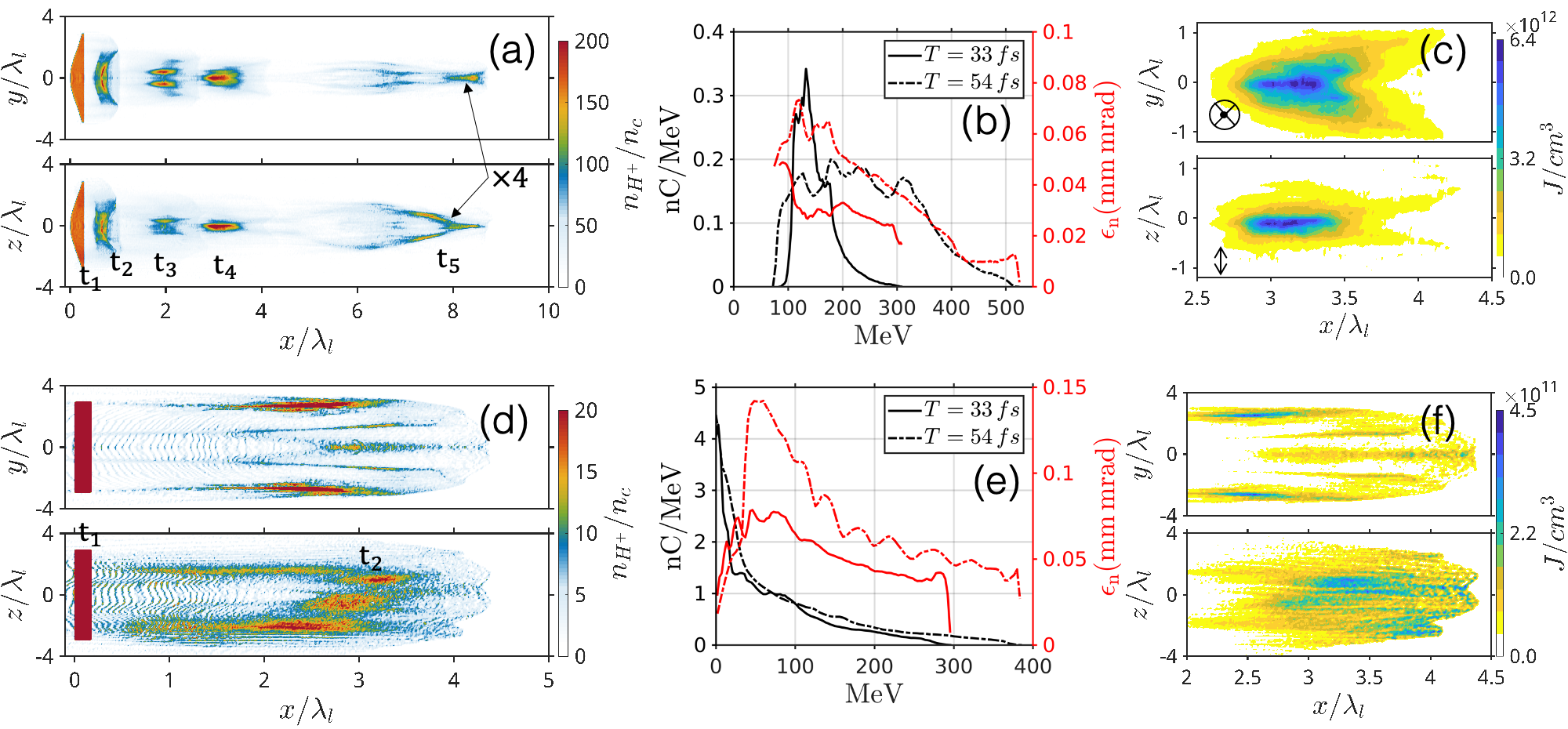} 
\caption{All figures are plotted for LP laser pulse with $\epsilon = 0.; a_0 = 113$.  Top Panel: Shaped Target.  (a) Time snapshots of ion density at $\mathrm{t_1 = 0} $, $\mathrm{t_2 = 14.5 \, fs} $, $\mathrm{t_3= 25 \, fs}$, $\mathrm{t_4 = 33 \, fs}$, and $\mathrm{t_6 = 54 \, fs} $. (b) Proton energy spectrum (black line; left scale) and emittance (red; right scale) at time instances $t = 33 \,  fs$ (solid) and $t =  54\,  fs$ (dash-dot).  (c) Time snapshots of ion density at $\mathrm{t_1 = 0} $,  and $\mathrm{t_2 = 33 \, fs}$. Bottom Panel: Planar Target.(d) Snapshot of ion densities.  (e) and (f) follow (b) and (c), respectively,  for the planar target.}
\label{Fig4}
\end{figure}

In the previous section, we investigated the influence of an elliptically polarized (EP) laser pulse on a shaped target, comparing results to those obtained with a planar target under the same laser conditions. This section examines the effects of a linearly polarized (LP) laser pulse on a shaped target. Figure \ref{Fig4} presents the outcomes for both shaped and planar targets under an LP laser pulse with a peak normalized vector potential of $a_0 = 112$. The central thickness of the shaped target has been optimized to $d_0 = 0.290 \, \mu m$ to balance radiation pressure and space charge forces, while for the planar target, it is set to $d_0 = 0.180 \, \mu m$. Figures \ref{Fig4}(a) and \ref{Fig4}(d) show the spatio-temporal evolution of the shaped and planar targets, respectively.

For the shaped target, the radial focusing force compresses the target over a shorter distance than observed in Case I, with the peak proton density reaching $n_{H^{+}} = 150 n_c$. This compression is illustrated in Fig. \ref{Fig4}(a), where maximum proton density is attained at $t_f$, reaching a peak value of $n_{H^{+}} = 150 n_c$ (hotspot). In both cases, target acceleration is driven by the non-oscillating component of the $\mathbf{J}  \times \mathbf{B}$ force from the LP pulse. The oscillating component of the $\mathbf{J} \times \mathbf{B}$ force, however, generates hot electrons with a temperature of $T_h \approx 40 \, \mathrm{MeV}$, which is approximately $20 \%$ lower than the ponderomotive scaling temperature, given by $T_h = (\gamma n_c / n_{0})^{1/2} \left[ \left(1 + a_0^2 \right)^{1/2} -1 \right] m_e c^2 \approx 50 \, \mathrm{MeV}$ \citep{Wilks_TNSA}. The discrepancy between simulated and theoretical temperatures can be attributed to the small laser spot size \citep{Dover_prl_2020}. Nonetheless, the generation of hot electrons significantly impacts the acceleration dynamics of the shaped target, as shown in Fig. \ref{Fig4}(a). The hot electrons penetrate the target quickly, forming a double-peaked electric field due to the non-oscillating (front surface) and oscillating (rear surface) components of the ponderomotive force \citep{Zhuo_prl_2010}. Owing to the surface curvature, the rear sheath electric field exerts a radially focusing force on the ions, which may contribute to a shorter focal length and focusing duration than initially designed, with protons reaching energies more than 100 MeV. The proton energy peaks at $E_k \approx 140\, \mathrm{MeV}$ (black solid line, Fig. \ref{Fig4}(b)), with a small normalized emittance of $\epsilon_n \simeq 0.027 \, \mathrm{mm\, mrad}$ (red solid line, Fig. \ref{Fig4}(b)) at the peak of the proton energy spectrum.

In contrast, the planar target undergoes considerable degradation due to both hot electron generation and radial expansion from the finite pulse width, as shown in Fig. \ref{Fig4}(d) at $t = 23.7 \, \mathrm{fs}$. Furthermore, Rayleigh-Taylor (RT) instabilities fragment the target in the transverse direction, leading to complete disruption. This instability causes the target to become transparent early in the simulation, effectively terminating RPA for the planar target, as seen in Fig. \ref{Fig4}(d) at $t = 46.2\, \mathrm{fs}$. As a result, the proton energy spectrum for the planar target is exponential, lacking a distinct energy peak, and the normalized emittance is high, around $0.145 \, \mathrm{mm \, mrad}$. By comparison, the shaped target achieves a peak proton energy density of approximately $6.4 \times 10^{12} \, \mathrm{J/cm^3}$, which is an order of magnitude higher than the $4.5 \times 10^{11} \, \mathrm{J/cm^3}$ obtained for the planar target, as shown in Fig. \ref{Fig4}(e).

\section{Discussions}
We have compared the outcomes of laser-target interactions using circularly, elliptically, and linearly polarized (CP, EP, and LP) laser pulses on both shaped and planar targets. We now turn to a comparative analysis of the shaped target under various polarization conditions. As discussed in Section II, the shaped target remains opaque under an EP laser pulse but is affected by hot electron generation. In Case I (shaped target; CP pulse), the proton energy peaks at $E_k \approx 230\, \mathrm{MeV}$, while in Case III (shaped target; EP pulse), the peak energy is $E_k \approx 190 \, \mathrm{MeV}$. To achieve energy comparable to a CP-driven shaped target using an EP-driven shaped target, we extended the laser-target interaction time. Given that the shaped (LILA) target behaves analogously to a material lens, it begins to defocus after $t_f \approx 46.2$, resulting in a reduction of accelerated ions over time. This effect is observable in Fig. \ref{Fig3}(a) at $t = 56.8$, where the peak density drops to $n_{H^{+}} \approx 60 n_c$ following its peak at $t_f = 46.2\, \mathrm{fs}$. Nevertheless, the peak of the proton energy spectrum (left scale; black dot-dashed line) reaches $E_k \approx 230\, \mathrm{MeV}$, matching the energy peak for the CP-driven target. Notably, beam quality declines and emittance increases over time, from $0.035 \, mm \, mrad$ (right scale; solid red line) to $0.06 \, mm \, mrad$ (right scale; dot-dashed red line). While hot electron generation from the EP pulse influences the RPA mechanism, a proton bunch from an EP laser-driven shaped target can still achieve energy comparable to that obtain a CP-driven shaped target, although remaining lower than the energy observed in the planar target case (Case IV). For comparison, we extended the simulation for the planar target, yet due to self-induced transparency, the planar target protons were not accelerated to significantly higher energies. A few ions did achieve increased cutoff energy as a result of relativistic induced transparency acceleration.

Similarly, an LP-driven target remains opaque at the time of focusing, allowing the laser-target interaction to continue. In Case V (shaped target; LP pulse), the peak proton energy is $E_k \approx 140 \,\mathrm{MeV}$. To achieve energy comparable to that of a CP-driven shaped target, we sustained the LP pulse interaction with the shaped target for an extended duration. Due to target defocusing, the target density decreases to $n_{H^+}/n_c \approx 50$ at $t \approx 54$ in Fig. \ref{Fig4}(a), and the emittance doubles from $\epsilon_n \simeq 0.027 \, mm \, mrad$ (right scale; solid red line) to $\epsilon_n \simeq 0.06 \, mm \, mrad$ (right scale; dot-dashed red line).

\begin{figure}
\centering
\includegraphics[width=1\linewidth]{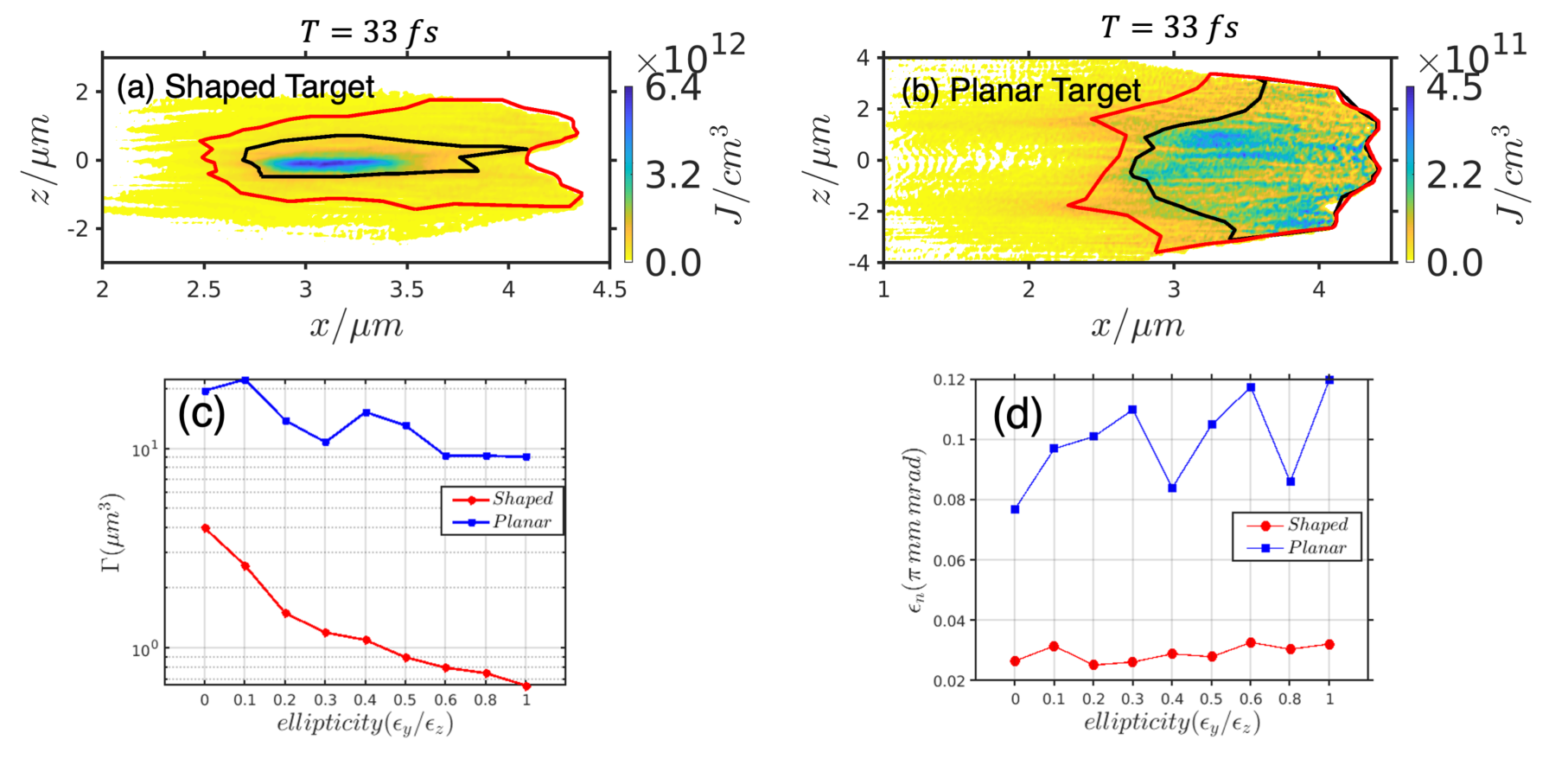} \label{fig6}
\caption{Top Panel: The snapshot of proton energy density in X-Z plane for LP laser pulse at the time $T = 33 \, fs$ for the (a) shaped, and the (b) planar target. The loop contains those protons, sum of whose energy is 3.5 J (inside black loop) and 7 J (inside red loop). Bottom Panel: Minimum volume containing protons whose total energy is 3.5 J for different ellipticity, and (d) The normalized emittance $\epsilon_n$ for different ellipticity at the time of maximum compressor for each value of ellipticity.}
\end{figure}

To illustrate the degree of ion beam focusing for LP pulses in shaped and planar targets, we present the ion kinetic energy density in Fig. \ref{fig6}(a) for the shaped target (Case V) and in Fig. \ref{fig6}(b) for the planar target. In these figures, red and blue closed loops enclose ions with total energies of $7 \, \mathrm{MeV}$ (red loop) and $3.5 \, \mathrm{MeV}$ (black loop), respectively. The loop volume differs significantly between the shaped and planar targets. For clarity, we plotted the volume containing ions with a total energy of $3.5\, \mathrm{MeV}$ for various values of laser polarization ellipticity in Fig. \ref{fig6}(c). It is evident that an LP-driven shaped target occupies a much smaller volume than a CP-driven planar target, and the volume difference increases as the polarization ellipticity increases, i.e., as the pulse transitions from LP to CP, the volume occupied by shaped and planar targets diverges further.

Another indicator of beam focusing quality is emittance. For comparison, we have plotted the normalized emittance across different laser polarization ellipticities in Fig. \ref{fig6}(d), where the red circle-dash line denotes emittance for the shaped target and the blue square-dash line for the planar target. Normalized emittance measurements were taken at the maximum compression point for shaped targets, with the same time instance used for planar targets. The emittance for the shaped target remains nearly constant across all ellipticity values and is consistently four times smaller than that of the planar target. 

\section{Summary}

We present the impact of various laser polarizations—circular (CP), elliptical (EP), and linear (LP)—on laser-driven ion acceleration in shaped and planar targets. By systematically comparing the outcomes for CP, EP, and LP pulses, we demonstrate the benefits of target shaping in optimizing proton acceleration, energy spectra, and beam emittance. Under CP pulses, shaped targets achieve peak proton energies up to $E_k \approx 230 \, \mathrm{MeV}$, significantly enhanced by target shaping effects that maintain beam density and reduce emittance. In contrast, EP and LP pulses yield somewhat lower energies but still provide favorable conditions for ion beam focusing, albeit with increased generation of hot electrons that impact the RPA mechanism. Notably, shaped targets display resilience against the onset of Rayleigh-Taylor instabilities and self-induced transparency, which otherwise degrade the performance of planar targets. Further analysis of the ion beam energy density and normalized emittance across different polarization modes reveals that shaped targets consistently outperform planar targets in energy density and maintain low emittance across all ellipticities. These findings highlight the efficacy of shaped targets focused ion beams with applications in advanced laser-driven acceleration schemes.

\bibliographystyle{jpp}

\bibliography{EP_LILA}

\end{document}